\documentclass[twocolumn,showpacs,superscriptaddress,preprintnumbers,amsmath,amssymb,prc]{revtex4-2}

\usepackage{graphicx}
\usepackage{dcolumn}
\usepackage{bm}
\usepackage{float}
\usepackage{color}
\usepackage{CJK}
\usepackage{ulem}
\usepackage[colorlinks,linkcolor=blue,urlcolor=blue,citecolor=blue]{hyperref}

\usepackage{tikz,xcolor,hyperref}

\definecolor{lime}{HTML}{A6CE39}
\DeclareRobustCommand{\orcidicon}{
	\begin{tikzpicture}
	\draw[lime, fill=lime] (0,0) 
	circle [radius=0.16] 
	node[white] {{\fontfamily{qag}\selectfont \tiny ID}};
	\draw[white, fill=white] (-0.0625,0.095) 
	circle [radius=0.007];
	\end{tikzpicture}
	\hspace{-2mm}
}
\foreach \x in {A, ..., Z}{%
	\expandafter\xdef\csname orcid\x\endcsname{\noexpand\href{https://orcid.org/\csname orcidauthor\x\endcsname}{\noexpand\orcidicon}}
}
\foreach \x in {A, ..., Z}{%
	\expandafter\xdef\csname orcid\x\endcsname{\noexpand\href{https://orcid.org/\csname orcidauthor\x\endcsname}{\noexpand\orcidicon}}
}

\begin{document}

\begin{CJK*}{UTF8}{gbsn}
\title{Information entropy for central $^{197}$Au+$^{197}$Au collisions 
with the ultrarelativistic quantum molecular dynamics model}

\author{X. G. Deng(邓先概)\orcidA{}}
\email[Corresponding author: ]{xiangai\_deng@fudan.edu.cn}
\affiliation{Key Laboratory of Nuclear Physics and Ion-beam Application (MOE), Institute of Modern Physics, Fudan University, Shanghai 200433, China}
\affiliation{Shanghai Research Center for Theoretical Nuclear Physics， NSFC and Fudan University, Shanghai 200438, China}

\author{Y. G. Ma(马余刚)\orcidB{}}
\email[Corresponding author: ]{mayugang@fudan.edu.cn}
\affiliation{Key Laboratory of Nuclear Physics and Ion-beam Application (MOE), Institute of Modern Physics, Fudan University, Shanghai 200433, China}
\affiliation{Shanghai Research Center for Theoretical Nuclear Physics， NSFC and Fudan University, Shanghai 200438, China}

\date{\today}

\begin{abstract}

This study investigates the multiplicity information entropy of hadrons, anti-hadrons, baryons, and net-protons in central \(^{197}\)Au+\(^{197}\)Au collisions with impact parameters of 0$-$3 fm, using the ultrarelativistic quantum molecular dynamics model (UrQMD) across various center-of-mass energies (\(\sqrt{s_{\rm NN}}\)) from 5.0 to 54.4 GeV. Our simulations employ hydrodynamic modes with different equations of state (EoS) and a default mode without hydrodynamics. The results reveal that the information entropies of baryons and net-protons are sensitive to the selected EoS. In particular, enhancements of the information entropies around \(\sqrt{s_{\rm NN}} \sim 30\) GeV, especially with chiral hadron gas and Bag model EoS,  indicate a phase transition or critical endpoint behavior. These findings highlight the importance of the EoS in understanding the thermodynamic properties of matter produced in high-energy collisions.

\end{abstract}


\maketitle

\section{Introduction}
\label{introduction}

In heavy-ion collisions, large nuclei, such as gold or lead, are smashed together at speeds approaching the speed of light. These extreme conditions produce such high temperatures and energy densities that protons and neutrons `melt', releasing quarks and gluons from their usual confinement. After hadronization, when quarks and gluons recombine into hadrons, a large number of particles are produced, offering valuable insights into the behavior of matter under extreme conditions. Experiments have shown that the multiplicity of particles produced in heavy-ion collisions is significantly higher than in proton-proton collisions, revealing the unique properties of heavy-ion collisions. 
Both the Relativistic Heavy Ion Collider and the Large Hadron Collider provide unique venues to study the properties of QGP as well as the quantum chromodynamics (QCD) phase diagram
\cite{PBM,Busza,AD20,Pandav,Chen,Shou}. The QCD phase diagram, which maps the relationship between temperature ($T$) and baryon chemical potential ($\mu_{B}$), has indeed been an ongoing area of research for several decades~\cite{CN75,GB02,KF08,KF11,LXF17}. As one of the primary goals of hot QCD physics~\cite{NT_Du,NT_Li,HeWB23,Ma23,ChenQ}, considerable effort has been devoted to locating the critical endpoint (CEP) in the QCD phase diagram using various methods and probes. 
For example, the ratio of viscosity to entropy density ($\eta/s$) \cite{DT03,SG06,NatPhys,DXG,DXG24} as well as measurements of the multiplicity and its cumulants, such as skewness and kurtosis \cite{PRL_QCD,MA11,HC18,JM19,STAR1_2021_netproton} provide tools to search for the CEP. In particular, fluctuations in conserved charges and baryon density~\cite{CS07,PRL_QCD,LiuC}, together with higher-order moments~\cite{MA09-1,JS12,Ko16,LF17,CaoRX}, and the combined ratio $N_{t}N_{p}/N_{d}^{2}$ of protons, deuterons, and tritons~\cite{KJ17,KJ18,STAR23,KoCM,SunKJ,Bleicher}, provide important insights into critical phenomena. 
In addition, the scaled factorial moments \cite{MaYG} of the identified charged hadrons 
exhibit a non-monotonic energy dependence in the 0-5\% most central collisions, with the scaling exponent reaching a minimum around $\sqrt{s_{NN}}$ = 27 GeV in \(^{197}\)Au + \(^{197}\)Au collisions, which could be related to the CEP \cite{STAR}. However, it is important to gain a theoretical understanding of the CEP from multiple perspectives.

In this work we focus on multiplicity distributions and their information entropies for (anti-)hadrons and (anti-)baryons as well as net-hadrons and net-baryons for Au + Au collisions from 5.0 GeV to 54.4 GeV/c with the ultrarelativistic quantum molecular dynamics (UrQMD) model \cite{Bleicher2}. Information entropies can be calculated basing on the multiplicity distributions, and their sensitivities to different equations of state (EoS) within hydrodynamic models can be studied. 

The concept of information entropy, first introduced by Shannon in the 20th century \cite{CE48}, has been widely adopted in various fields. In nuclear physics, the notion of multiplicity information entropy ($S$) was first proposed by Ma in 1999 and has been used in heavy-ion collisions to study the liquid-gas phase transition of nuclear matter~\cite{YGM99}. Detailed formulation of $S$ will be given in the next section.  Here we extend its application to relativistic heavy-ion collisions, in particular for the first time to anti- and net- hadrons (baryons). This may provide new insights into the properties of QGP and the QGP phase transition.

 From a theoretical perspective, the relationship between the multiplicity information entropy and the equation of state of nuclear matter arises from how particle multiplicity and its fluctuation reflects the thermodynamic properties of nuclear matter. 
Multiplicity, which represents the number of particles produced in a heavy-ion collision, is directly related to the energy density, temperature and phase of nuclear matter, all of which are described by EoS. In addition, fluctuations in particle multiplicity are expected to enhance near the phase transition regions and the critical point of nuclear matter, and the behavior of fluctuations is governed by the compressibility and specific heat of the system, which are encoded in the EoS.
The Shannon entropy derived from the multiplicity distributions of particles could be a powerful tool for studying these relationships. Under general consideration, Shannon entropy is sensitive to the following ingredients, e.g. (1) the phase of nuclear matter: lower in the hadronic phase, higher in the QGP phase, (2) phase transitions: peaks near the critical point due to enhanced fluctuations; (3) system size and collision energy: scaling with $s⋅V $, in which $s$ is entropy density and $V$ is system volume, reflects the underlying thermodynamic state dictated by the EoS.

Motivated by the above arguments, the relationship between the Shannon entropy and the equation of state of nuclear matter is a very interesting question, and this constitutes our main discussion in this article. The paper is organized as follows. Sect. II provides a brief overview of the UrQMD model and the concept of information entropy. Sect. III presents the calculations of the information entropies of hadrons and anti-hadrons, baryons and anti-baryons, as well as net-baryons and net-protons in Au + Au collisions at energies from 5.0 to 54.4 GeV/c using various potential parameter sets within the UrQMD framework. Multiplicity distributions, time evolution,  and energy dependence of the information entropies of different hadrons are also presented. Additionally, we compare the calculated net proton information entropy with experimental results and discuss their physical implications. Finally, a summary is given.

\begin{figure}[htb]
\setlength{\abovecaptionskip}{0pt}
\setlength{\belowcaptionskip}{8pt}
\includegraphics[scale=1.05]{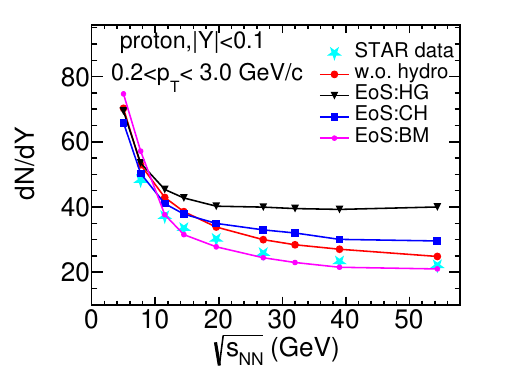}
\caption{Proton number in a unit of rapidity as a function of center of mass energy in central $^{197}$Au + $^{197}$Au collisions at impact parameter 0$-$3 fm for the rapidity $|Y|< 0.1$ and transverse momentum 0.2 $<  p_{T}< $3.0 GeV/c with different potential parameter modes of UrQMD. The STAR data is taken from Ref.~\cite{LH20}.}
\label{fig:fig1}
\end{figure}

\begin{figure*}[t]
\setlength{\abovecaptionskip}{0pt}
\setlength{\belowcaptionskip}{8pt}
\includegraphics[scale=0.84]{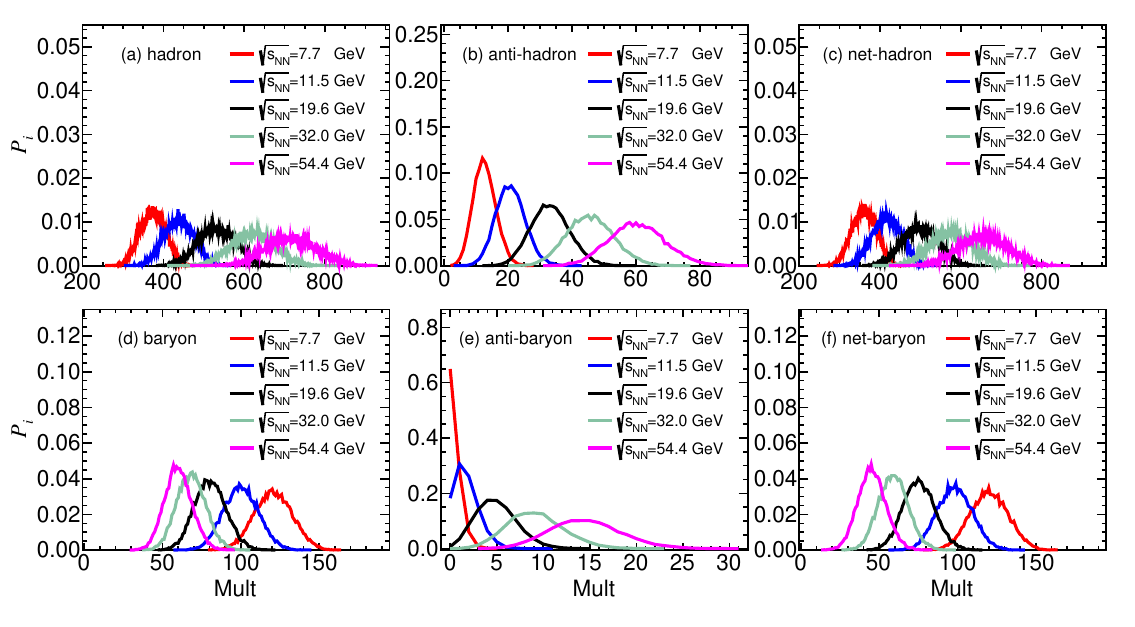}
\caption{Normalized multiplicity distribution of (a) hadrons, (b) anti-hadrons, (c) net-hadrons, (d) baryons, (e) anti-baryons, and (f) net-baryons at different center of mass energies in central $^{197}$Au + $^{197}$Au collisions at impact parameter 0$-$3 fm for $|Y|< 0.5$ and 0.2 $< p_{T}< $3.0 GeV/c with the standard mode UrQMD.}
\label{fig:fig2}
\end{figure*}

\begin{figure*}[t]
\setlength{\abovecaptionskip}{0pt}
\setlength{\belowcaptionskip}{8pt}
\centerline{
\includegraphics[scale=0.9]{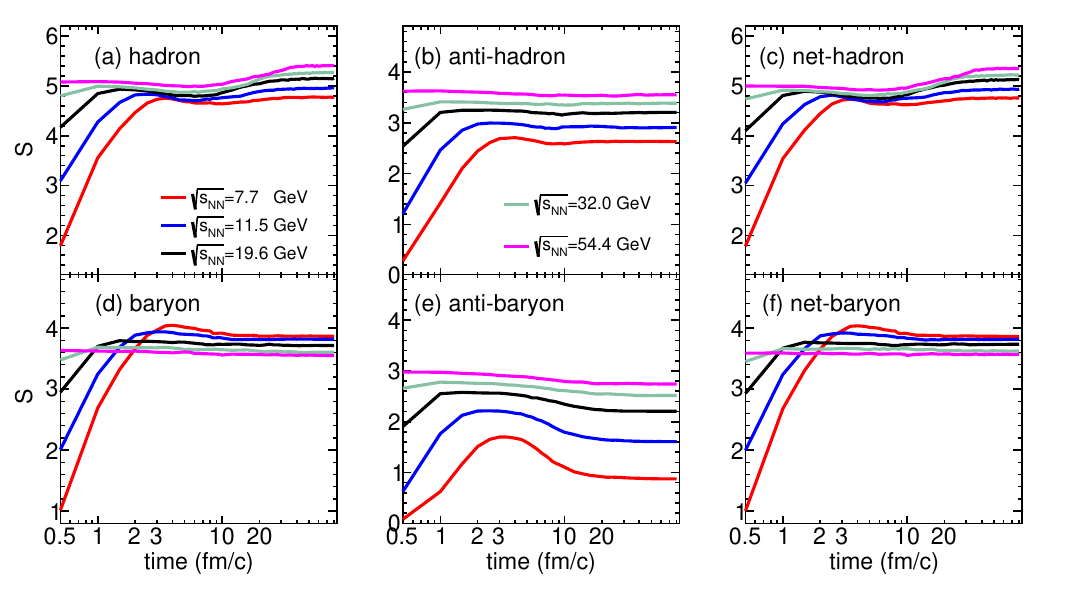}}
\caption{ Same as Fig.~\ref{fig:fig2} but for the time evolution of information entropies.}
\label{fig:fig3}
\end{figure*}

\begin{figure}[htb]
\setlength{\abovecaptionskip}{0pt}
\setlength{\belowcaptionskip}{8pt}
\includegraphics[scale=1.02]{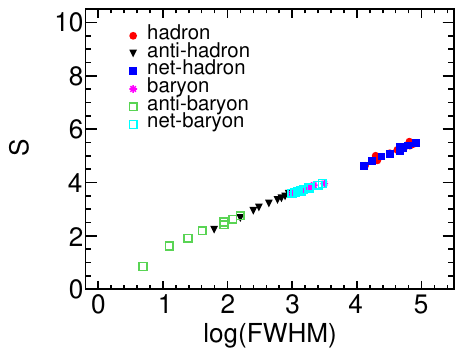}
\caption{Information entropies of (a) hadrons and (b)anti-hadrons as a function of log(FWHM) (FWHM: full width at half maximum) of their multiplicity distributions in central $^{197}$Au+$^{197}$Au collisions at impact parameter 0$-$3 fm with the standard mode UrQMD.}
\label{fig:fig4}
\end{figure}

\section{UrQMD model and information entropy}
\label{entropyandmodel}

One of the purposes of the transport model for relativistic heavy-ion collisions is to provide an effective solution of the relativistic Boltzmann equation \cite{SR80,HP08}
\begin{equation}
p^{\mu}\cdot \partial_{\mu} f_{i}(x^{\nu},p^{\nu})= \mathcal{C}_{i},                     
\label{ReBolzEq}
\end{equation}
where $f_{i}$ is the distribution function of particle species $i$ and $\mathcal{C}_{i}$ is the collision term. In the present work, simulations are based on  the framework of the Ultrarelativistic Quantum Molecular Dynamics model (UrQMD), which includes about 60 baryonic species and 40 mesonic species, as well as their anti-particles \cite{SA98,MB99,HP08,JS18,QiaoFH}. The model includes particle rescattering, color string fragmentation, and the formation and decay of hadronic resonances. UrQMD is one of  successful transport models for describing heavy-ion collisions across various energy ranges, spanning from the BNL Alternating Gradient Synchrotron (AGS) energies ($E_{\text{lab}}$ = 1 $-$10 $A$GeV) to SPS energies ($E_{\text{lab}}$ = 20 $-$160 $A$GeV), and extending to the full RHIC energy ($\sqrt{s_{\text{NN}}}$ = 200 GeV) and the energy levels of the CERN Large Hadron Collider (up to 2.76 TeV for Pb+Pb collisions)~\cite{MM09,PPB10,SS19,LiQF_NT,QiaoFH,LiQF_NT2,LiQF_NST,SSS_NST,Li_SCPMA}.

It should be noted that for the standard mode of UrQMD,  the hadrons and strings cannot be effectively modeled at higher energy, such as $\sqrt{s_{\rm NN}}$ $>$ 100 GeV. 
Thus, one utilizes a 3D$+$1D ideal fluid dynamics to describe the initial stage of the collision. As the system cools and reaches freeze-out conditions, one can transform the fluid dynamical fields to discrete hadrons using the Cooper-Frye equation,
\begin{equation}
E\frac{d^{3}N_{i}}{dp^{3}} = \int_{\Gamma} f_{i}(x^{\nu},p^{\nu}) p^{\nu}d\Gamma_{\nu},                     
\label{COOPEREq}
\end{equation}
Here $f_{i}(x^{\nu},p^{\nu})$ is the grand canonical Bose- or Fermi-distribution function for particle species, which depends on the local temperature $T(x)$ and chemical potentials $\mu_{i}$(x) \cite{JS18}. For hadron propagation, it follows the default mode of operation. The entire process, combining micro-Boltzmann transport and macro-hydrodynamics, is referred to as the hybrid mode. In hydrodynamics, the EoS as a function of energy and the baryon number density is needed as an additional input to calculate the pressure, temperature, and chemical potential \cite{HP08}. Three kinds of EoS, namely hadron gas (HG) EoS without deconfinement transition (EoS:HG) \cite{DZ02}, chiral+hadron gas (CH) EoS with first order transition and critical endpoint (EoS:CH)  \cite{JS11}, and bag model (BM) EoS with a strong first-order phase transition between QGP and hadronic phase (EoS:BM)  \cite{DH95}, are employed in our simulations \cite{HP08,LI2009111,YB16}. As a comparison, except for the hybrid mode, the standard (or default) mode is considered here. In UrQMD with hydrodynamics, as the system evolves and cools down, a particlization procedure is needed, which is dealt with in Cooper-Frye formalism. In this work, the default scenario known as the gradual particlization scenario (GF) is applied to the particlization. For the check, the proton number per rapidity as a function of center of mass energy is given in Fig.~\ref{fig:fig1}. The model results with different modes of UrQMD are compared to the data \cite{LH20}. As shown by magenta line from EoS mode with BM, the values are the closest to the STAR data~\cite{LH20}, and while red line which is the result of the default mode  is also close to the STAR data. But the result with CH EoS is higher than the data when $\sqrt{s_{\rm NN}}$ is above 20 GeV, and the HG EoS trend diverges beyond 10 GeV.

With the help of UrQMD, one can extract the multiplicity information entropy, which was initially proposed by Ma in Ref.~\cite{YGM99}, defined as:
\begin{equation}
S = -\sum_{i} P_{i}\ln P_{i},                    
\label{InEntropy}
\end{equation}
where $i$ is the total number of  all particles or a specific type of particle (e.g. hadrons or baryons etc.) produced in an event and $P_{i}$ = $N_{i}/N_{t}$ is the normalized multiplicity  probability. Here $N_{i}$ and $N_{t}$ denote event number of the multiplicity `$i$' and total event number, respectively and the bin width is set to `1'. It should be mentioned that $\sum_{i}P_{i}$ = 1. Mathematically, this entropy $S$ is called Shannon entropy~\cite{CE48}. Some applications on multiplicity information entropy have been presented in the field \cite{CWM18,LF20,PuJie23,PRD,PLB,WeiHL}. 
As discussed in previous section, the  information entropy should be related to the EOS. We should note that even though here  we just take information entropy as an observable of the microscopic complexity and disorder of a system, we should point out that other kinds of entropy, such as the Tsallis entropy \cite{Tsallis} and the Renyi entropy \cite{Renyi} etc., essentially follows the similar results since those information entropies  are all measures of uncertainty or disorder in a system, and they generalize the concept of entropy in different ways. These three entropies are closely related, but each emphasizes different aspects of the system's distribution.

In the present work, we extract information entropy in central $^{197}$Au+$^{197}$Au collisions at impact parameter 0$-$3 fm with transverse momentum cut $0.2 < p_{T} < 3.0$ GeV/c and rapidity cut $|$Y$|$$<$0.5.

\section{Results and discussion}
\label{resultsSH}

Figure~\ref{fig:fig2} illustrates the normalized multiplicity distribution of (a) hadrons, (b) anti-hadrons, (c) net-hadrons, (d) baryons, (e) anti-baryons, and (f) net-baryons at different center-of-mass energies in a framework of standard mode of UrQMD. It should be noted here that the hadrons and baryons denote positive particles. As shown in the figure, the production of hadrons (or anti-hadrons or net-hadrons) increases with collision energy, and their distributions broaden with increasing energy. However, for baryons and net-baryons, the opposite trend is observed, with narrower distributions at higher energies. The multiplicity distribution of anti-baryons follows the  similar behavior as that of hadrons.

One can easily obtain information entropy from these multiplicity distributions in Fig.~\ref{fig:fig2}. As shown in Fig.~\ref{fig:fig3}, we present the time evolution of information entropies for different particle species. The information entropies of hadrons, anti-hadrons, net-hadrons, baryons, anti-baryons, and net-baryons all increase with time, reaching more than 90\% of the total entropies within 0$-$5 fm/c. It indicates that the system reaches the chemical freeze-out fast within 0$-$5 fm/c. Moreover, the information entropy for hadrons is higher than that for anti-hadrons, and the similar behavior is observed between baryons and anti-baryons. This could be due to greater disorder or fluctuation for positive baryons and hadrons, as their abundant production induces greater fluctuations.

\begin{figure}[htb]
\setlength{\abovecaptionskip}{0pt}
\setlength{\belowcaptionskip}{8pt}
\includegraphics[scale=1.0]{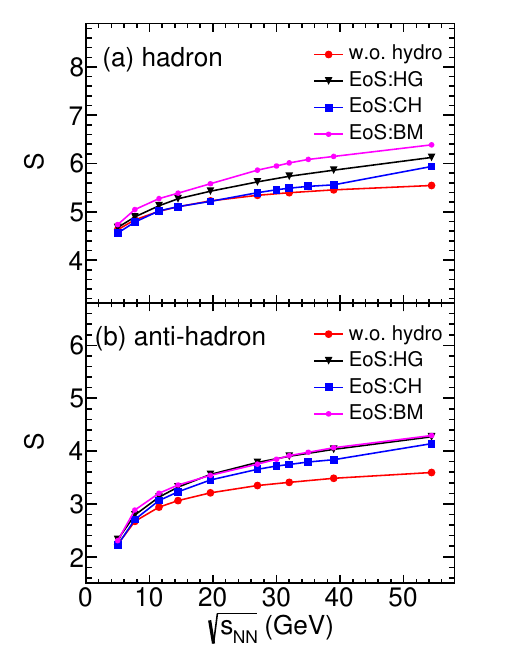}
\caption{Information entropies of (a) hadrons and (b) anti-hadrons as a function of center of mass energy in central $^{197}$Au+$^{197}$Au collisions at impact parameter 0$-$3 fm.}
\label{fig:fig5}
\end{figure}

\begin{figure}[htb]
\setlength{\abovecaptionskip}{0pt}
\setlength{\belowcaptionskip}{8pt}
\includegraphics[scale=1.0]{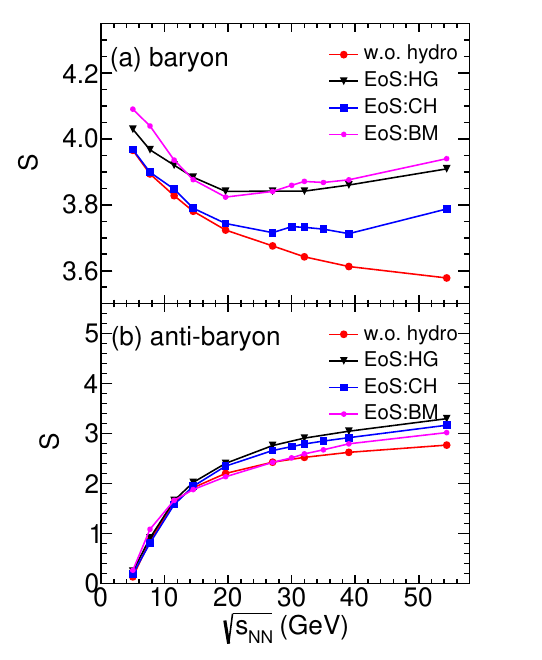}
\caption{Information entropies of (a) baryons and (b)anti-baryons as a function of center of mass energy in central $^{197}$Au+$^{197}$Au collisions at impact parameter 0$-$3 fm.}
\label{fig:fig6}
\end{figure}

\begin{figure}[htb]
\setlength{\abovecaptionskip}{0pt}
\setlength{\belowcaptionskip}{8pt}
\includegraphics[scale=1.0]{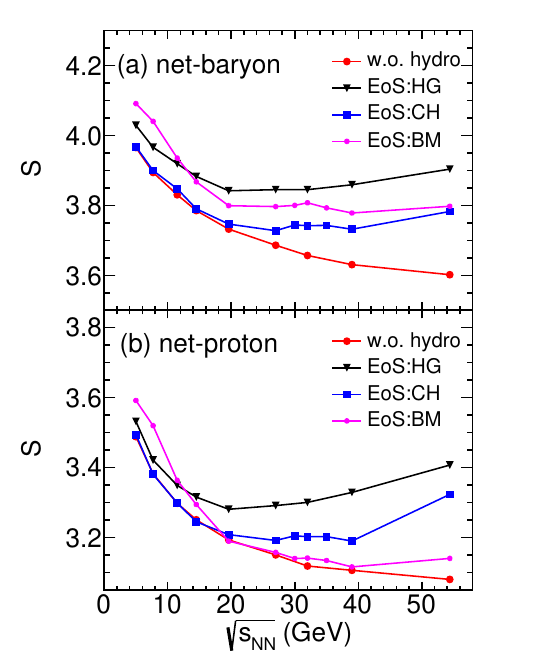}
\caption{Information entropies of (a) net-baryons and (b) net-protons as a function of center of mass energy in central $^{197}$Au+$^{197}$Au collisions at impact parameter 0$-$3 fm. }
\label{fig:fig7}
\end{figure}

We can obtain the information entropy at the final stage and establish the relation between information entropy and center of mass energy. Although information entropy is calculated from the multiplicity probability $P_{i}$ as shown in Eq.~(\ref{InEntropy}), it is observed that information entropy is closely linked to the width of the multiplicity probability distribution. For instance, Fig.~\ref{fig:fig4} illustrates the relation between information entropy and the logarithm of the full width at half maximum (FWHM) of $P_{i}$ with the standard UrQMD mode, revealing an approximately linear relationship. Actually, if the multiplicity distribution is a perfect Gaussian distribution with FWHM, the information entropy can be mathematically deduced by the FWHM, i.e. $S = const +  ln(FWHM)$. This suggests that entropy really reflects the fluctuation or disorder of the system through the multiplicity distribution; specifically, a wider multiplicity probability distribution corresponds to higher information entropy.  As the width increases, the shape of the multiplicity distribution becomes flatter and more spread out. In an extreme case  when the maximum entropy $S_{max}$ is reached, all corresponding probabilities $P_{i}$ of certain multiplicities $i$ tend to the same. 
From the point of view of $P_i$ fluctuations, there is no uncertainty at $S_{\text{max}}$. This means that all multiplicities $i$ have the same probability $P_i$.

In our simulations, we explore four scenarios: the standard mode without hydrodynamics and three hydrodynamic modes employing different Equations of State (EoS): EoS:HG, EoS:CH, and EoS:BM, as shown in Fig.~\ref{fig:fig5}. In both of hadrons and anti-hadrons, information entropies increase with collision energy. It stems from that both the FWHMs of multiplicity distributions of hadrons and anti-hadrons increase as collision energy, which illustrates that increasing uncertainty for predicting how many hadrons' multiplicity in a certain collision event.

In Fig.~\ref{fig:fig5} (a), a comparison between these four cases is present, revealing that the information entropy is  the  lowest for the scenario without hydrodynamics.  This  means that the FWHM for the w.o. hydro mode is smaller and indicates that the hydro modes make the fluctuation larger since that for a distribution, the wider FWHM means a larger standard deviation (or fluctuation). And the hydro mode with the EoS:BM is the largest one. However, the information entropy of the anti-hadrons with EoS:BM becomes close to the other two EoSs. If we compare the sensitivity of the information entropy (Fig.~\ref{fig:fig5}) or proton number (Fig.~\ref{fig:fig1}) to the UrQMD parameter set, we can say the information entropy is very sensitive to the hydo potential or not. For more details, Fig.~\ref{fig:fig1} shows a very close proton number for the cases EoS:BM and w.o. hydro, but Fig.~\ref{fig:fig5} shows a large difference of information entropy for the same cases.

Further, we give baryon and anti-baryon information entropies, as shown in Fig.~\ref{fig:fig6}. One should notice that the behavior of baryons is different from the anti-baryons. For the latter with increasing collision energy, the information entropy increases monotonically as hadrons and anti-hadrons do. The information entropies of baryons, however, for all hydro cases with EoS exhibit valley shapes with a minimum at around 20 GeV, 
and consistent with a previous study utilizing a multiphase transport (AMPT) model in Ref.~\cite{PuJie23}. It indicates that the introduction of  the  hydro mode induces an additional fluctuation starting from $\sim$ 20 GeV. 

Furthermore, in comparison with different EoS cases, one observes an enhancement of the information entropy around $\sqrt{s_{\rm NN}} = 30$ GeV for both the EoS:CH and EoS:BM cases, which include first-order phase transitions. As expected in relativistic heavy-ion collisions, large fluctuations in the baryon density would arise at energies where the system crosses   the first-order phase boundary or approaches to the CEP~\cite{AD20}. From the point of view of information entropy, large fluctuations in the baryon density could lead to broader multiplicity probability distribution, resulting in higher information entropy. In other words, the enhancement around 30 GeV collision energy, together with the non-monotonic behavior, could correspond to the first-order phase boundary or CEP. For the anti-baryons, however, their information entropies only show a monotonic increase as collision energy, as shown in Fig.~\ref{fig:fig6} (b).

\begin{figure}[htb]
\setlength{\abovecaptionskip}{0pt}
\setlength{\belowcaptionskip}{8pt}
\includegraphics[scale=1.0]{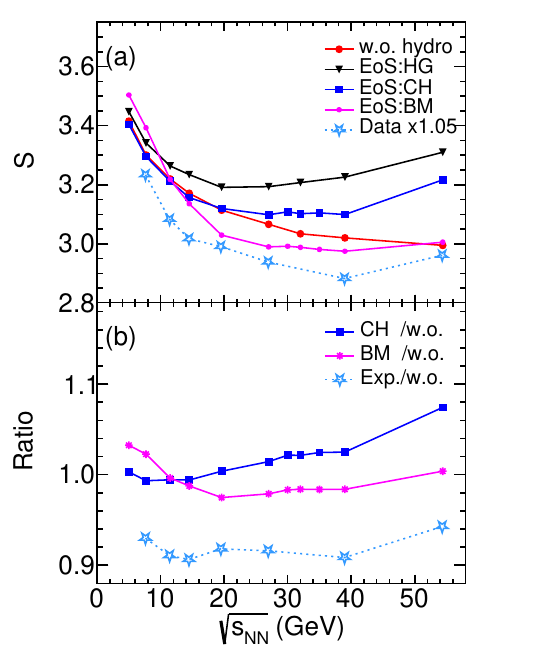}
\caption{Net-proton information entropies ((a), same as Fig.~\ref{fig:fig7}(b) but for the $p_{T}$ cut) and ratios of the net-proton information entropy of EoS:CH, of EoS:BM, and of the STAR data to that w.o. hydro  (b) as a function of center of mass energy in central $^{197}$Au+$^{197}$Au collisions at impact parameter 0$-$3 fm for $|Y|< 0.5$ and 0.4 $< p_{T}< $2.0. The experimental data points are  deduced from the net-proton distributions in Ref.~\cite{STAR1_2021_netproton}.}
\label{fig:fig8}
\end{figure}

Now let us turn to information entropies of net-baryons or net-protons. In previous works, fluctuations of the net-baryon number and net-proton number have been extensively discussed for the CEP in the point of view of higher order moments \cite{KM15,CH16,LXF17}. From the STAR measurement, the $\kappa \sigma^{2}$, where $\kappa$ and $\sigma$ represent kurtosis and variance respectively, of net-protons shows a nonmonotonic energy dependence, i.e. there appears a minimum $\kappa \sigma^{2}$ which is close to zero around 20 GeV at 0-5$\%$ of most central Au+Au collisions~\cite{STAR1_2021}. The non-monotonic 
$\kappa \sigma^{2}$  versus collision energy is qualitatively consistent with expectations from the QCD-based model expectation that, the higher the order of the moments is, the more sensitive it is to physics processes such as a critical point  \cite{PRL_QCD}, or maybe  an indication of onset the hydrodynamics process from the present work.
In addition, as mentioned before, enhancements of  the ratio $N_{t}N_{p}/N_{d}^{2}$ relative to the coalescence baseline as well as a minimum scale exponent of  intermittency at the STAR BES experiments are also observed in the $0\%-10\%$ most central collisions at around 27 GeV~\cite{STAR23}.
Here the information entropies for multiplicity distributions of net-baryons and net-protons are investigated and displayed in Fig.~\ref{fig:fig7}. 
From Fig.~\ref{fig:fig7}(a), one can see that information entropy with EoS:HG shows smooth line as the result of baryons. However, both EoS:BM and EoS:CH exhibit broad enhancements around 30 GeV, indicating possible connections to the first-order phase transitions or critical phenomena.

To compare with the experimental data, we fine-tuned the $p_{T}$ range as depicted in Figure \ref{fig:fig8}. Fig.~\ref{fig:fig8}(a) displays the net-proton information entropy as Fig.~\ref{fig:fig7}(b) but with 0.4 $< p_{T}< $2.0, alongside the experimental data extracted from net-proton distributions reported in Ref.~\cite{STAR1_2021_netproton}. Figure~\ref{fig:fig8}(b) shows the ratios of net-proton information entropy for EoS:CH, EoS:BM, and the STAR data relative to the results without hydrodynamics (denoted as CH/w.o., BM/w.o., and Exp./w.o., respectively), where the entropy without hydrodynamics serves as the baseline. The experimental data shows an enhancement around 20 GeV compared to the baseline entropy (without hydrodynamics), coinciding with the minimum $\kappa \sigma^2$ value reported in Ref.~\cite{STAR1_2021}. Meanwhile, the UrQMD simulations with EoS:BM and EoS:CH also exhibit slight enhancements, although these appear at higher collision energies, around 30 GeV. This difference suggests that information entropy could serve as an alternative observable for the QGP phase transition.

\section{Conclusions}
\label{summary}

In summary, we simulated central $^{197}$Au+$^{197}$Au collisions using the UrQMD model considering both a standard mode and hybrid modes with three types of EoSs, i.e. EoS:HG without deconfinement transition, EoS:CH with the first-order phase transition and critical endpoint, and EoS:BM with a strong first-order phase transition between the QGP and hadronic phases. By calculating the multiplicity information entropies of hadrons, anti-hadrons, baryons, anti-baryons, net-baryons and net-protons, we investigate the possible phase transition and EoS effects for the hydrodynamics. The dependence of information entropy on collision energy illustrates the relationship between the width of the multiplicity probability distribution of particles and collision energy. With the help of baryon and net-baryon information entropies, it is observed that both exhibit valleys around $\sqrt{s_{\rm NN}} \approx 30$ GeV in the hydro hybrid modes, signaling the onset of hydrodynamic behavior in the model. Additionally, an enhancement emerges around $\sqrt{s_{\rm NN}}$ = 30 GeV with EoS:CH and EoS:BM  which is consistent with recent experimental observations from $N_{t}N_{p}/N_{d}^{2}$ as well as intermittency analysis which give enhancement around $\sqrt{s_{\rm NN}}$=27 GeV, that may suggest the CEP. In one word, the present work suggests that the information entropy could serve as an alternative observable on the QGP phase transition.

\begin{acknowledgments}
Authors thank Dr. Chen Zhong for maintaining the high-quality performance of Fudan supercomputing platform for nuclear physics and the discussions with Dr. Kai-Jia Sun. This work received partial support from the National Natural Science Foundation of China under Contract Nos.\ $12205049$, $12347149$, $11890714$, and $12147101$，$11925502$, the Strategic Priority Research Program of CAS under Grant No. XDB34000000.
\end{acknowledgments}

\end{CJK*}


\begin{thebibliography}{99}

\bibitem{PBM}
P. Braun-Munzinger, V.  Koch, T. Sch\"afer, J. Stachel, \href{https://doi.org/10.1016/j.physrep.2015.12.003}{Phys. Rep. {\bf 621}, 76 (2016)}.


\bibitem{Busza}
W. Busza, K. Rajagopal, W.  van der Schee, \href{https://doi.org/10.1146/annurev-nucl-101917-020852}{Ann. Rev. Nucl. Part.  Sci. {\bf 68}, 339 (2018)}.

 \bibitem{AD20}  A. Bzdak,  S. Esumi, V. Koch, J. F.  Liao, M. Stephanov, N.  Xu, 
\href{https://doi.org/10.1016/j.physrep.2020.01.005} {Phys. Rep. {\bf 853}, 1 (2020)}.  

\bibitem{Pandav}
A.  Pandav, D. Mallick, B.  Mohanty, \href{https://doi.org/10.1016/j.ppnp.2022.103960}{Prog. Part. Nucl. Phys. {\bf{125}}, 103960 (2022)}.

\bibitem{Chen}
J. H. Chen, X. Dong, X. He  {\it et al.}, \href{https://doi.org/10.1007/s41365-024-01591-2 }{ Nucl. Sci. Tech. {\bf 25}, 214 (2024)}.

\bibitem{Shou}
Q. Y. Shou, Y. G. Ma,  S. Zhang {\it et al.}, \href{https://doi.org/10.48550/arXiv.2409.17964}{arXiv:2409.17964}.

\bibitem{CN75} N. Cabibbo and G. Parisi, \href{https://doi.org/10.1016/0370-2693(75)90158-6}{Phys. Lett. B {\bf 59}, 67 (1975)}.
\bibitem{GB02} G. Baym, \href{https://doi.org/10.1016/S0375-9474(01)01342-2}{Nucl. Phys. A {\bf 698}, xxiii-xxxii (2002)}.
\bibitem{KF08} K. Fukushima, \href{https://doi.org/10.1088/0954-3899/35/10/104020}{J. Phys. G: Nucl. Part. Phys. {\bf 35}, 104020 (2008)}.
\bibitem{KF11} K. Fukushima and T. Hatsuda, \href{https://doi.org/10.1088/0034-4885/74/1/014001}{Rep. Prog. Phys. {\bf 74},   014001 (2011)}. 
\bibitem{LXF17} X. F. Luo and N. Xu, 
 \href{https://doi.org/10.1007/s41365-017-0257-0}{Nucl. Sci. Tech. {\bf 28}, 112 (2017)}.




\bibitem{NT_Du}
Y. L. Du, C. M. Li, C. Shi, et al., 
\href{https://doi.org/10.11889/j.0253-3219.2023.hjs.46.040009}{Nucl. Tech. (in Chinese) {\bf 46}, 040009 (2023)}.


\bibitem{NT_Li}F. P. Li, L. G. Pang, X. N. Wang, 
\href{https://doi.org/10.11889/j.0253-3219.2023.hjs.46.040014}{Nucl. Tech. (in Chinese) {\bf 46}, 040014 (2023) }.

\bibitem{HeWB23}
Wan-Bing He, Yu-Gang Ma, Long-Gang Pang, Hui-Chao Song, Kai Zhou, \href{https://doi.org/10.1007/s41365-023-01233-z }{Nucl. Sci. Tech. {\bf 34}, 88 (2023)}.

\bibitem{Ma23}
Yu-Gang Ma, Long-Gang Pang, Rui Wang, Kai Zhou, 
\href{https://doi.org/10.1088/0256-307X/40/12/122101 }{Chin. Phys. Lett. {\bf 40}, 122101 (2023) }.

\bibitem{ChenQ}Qian Chen, Guo-Liang Ma and Jin-Hui Chen,
\href{https://doi.org/10.11889/j.0253-3219.2023.hjs.46.040013}{Nucl. Tech. (in Chinese), {\bf 46}, 040013 (2023).} 





\bibitem{DT03} D. Teaney, \href{https://doi.org/10.1103/PhysRevC.68.034913}{Phys. Rev. C {\bf 68}, 034913 (2003)}.

\bibitem{SG06} S. Gavin and M. Abdel-Aziz, \href{http://dx.doi.org/10.1103/PhysRevLett.97.162302}{Phys. Rev. Lett. {\bf 97}, 162302 (2006)}.

 \bibitem{NatPhys} J. E. Bernhard, J. S. Moreland, S. A. Bass, \href{http://dx.doi.org/10.1038/s41567-019-0611-8}{Nat. Phys. {\bf 15}, 1113-1117 (2019)}. 

\bibitem{DXG} X. G. Deng,  P. Danielewicz, Y. G. Ma, H. Lin,  Y. X. Zhang, \href{https://doi.org/10.1103/PhysRevC.105.064613}{Phys. Rev. C {\bf 105}, 064613 (2022)}.

 \bibitem{DXG24} X. G. Deng, D. Q. Fang, Y. G. Ma, \href{https://doi.org/10.1016/j.ppnp.2023.104095}{Prog. Part. Nucl.  Phys., {\bf 136}, 104095 (2024)}.

  \bibitem{PRL_QCD}
M. A. Stephanov, \href{https://doi.org/10.1103/PhysRevLett.102.032301}{Phys. Rev. Lett. {\bf 102}, 032301 (2009)}.
 
\bibitem{MA11} M. A. Stephanov, \href{https://doi.org/10.1103/PhysRevLett.107.052301}{Phys. Rev. Lett. {\bf 107}, 052301 (2011)}.

\bibitem{HC18} C. Herold, M. Bleicher, M. Nahrgang {\it et al}., \href{https://doi.org/10.1140/epja/i2018-12438-1}{Eur. Phys. J. A {\bf 54}, 19 (2018)}.

\bibitem{JM19} J. M. Torres-Rincon and E. Shuryak,
\href{https://doi.org/10.22323/1.347.0176}
{Proceedings of Science {\bf 347}, 176 (2019)}.

              
\bibitem{STAR1_2021_netproton} J. Adam  {\it et al.} (STAR Collaboration),
\href{https://doi.org/10.1103/PhysRevLett.126.092301}{Phys. Rev. Lett. {\bf 126}, 092301 (2021)}.

              
   
 \bibitem{CS07} C. Sasaki, B. Friman, and K. Redlich, 
                        \href{https://doi.org/10.1103/PhysRevLett.99.232301}{Phys. Rev. Lett. {\bf 99}, 232301 (2007)}.  
   \href{https://doi.org/10.1103/PhysRevLett.102.032301}{Phys. Rev. Lett. {\bf102}, 032301 (2009)}. 


\bibitem{LiuC}C. Liu, X. G. Deng, Y. G. Ma, 
\href{doi: 10.1007/s41365-022-01040-y}{Nucl. Sci. Tech. {\bf 33}, 52 (2022). }


\bibitem{MA09-1} M. A. Stephanov, S. Ejiri, and M. Kitazawa,
\href{https://doi.org/10.1103/PhysRevLett.103.262301}{Phys. Rev. Lett. {\bf 103}, 262301(2009)}.



\bibitem{JS12} J. Steinheimer and J. Randrup,
\href{https://doi.org/10.1103/PhysRevLett.109.212301}{Phys. Rev. Lett. {\bf 109}, 212301 (2012)}.
        
\bibitem{Ko16} C. M. Ko and F. Li, \href{https://doi.org/10.1007/s41365-016-0141-3}{Nucl. Sci. Tech. {\bf 27}, 140 (2016).}
                    
\bibitem{LF17} F. Li and C. M. Ko,
\href{https://doi.org/10.1103/PhysRevC.95.055203}{Phys. Rev. C {\bf 95}, 055203 (2017)}.

\bibitem{CaoRX}
Ru-Xin Cao, Song Zhang, Yu-Gang Ma, \href{https://doi.org/10.1103/PhysRevC.106.014910}{Phys. Rev. C {\bf 106}, 014910 (2022)}.
                                       
\bibitem{KJ17} K. J. Sun, L. W. Chen, C. M. Ko, Z. Xu, 
 \href{https://doi.org/10.1016/j.physletb.2017.09.056}{Phys. Lett. B {\bf 774}, 103  (2017)}.  


 \bibitem{KJ18} K. J. Sun,  L. W. Chen, C. M. Ko, J. Pu, Z. Xu, 
 \href{https://doi.org/10.1016/j.physletb.2018.04.035}{ Phys. Lett. B {\bf 781}, 499 (2018).}

\bibitem{STAR23} M. I. Abdulhamid {\it et al.} (STAR Collaboration), 
\href{https://doi.org/10.1103/PhysRevLett.130.202301}{Phys. Rev. Lett. {\bf 130}, 202301 (2023)}.
             
\bibitem{KoCM} C. M. Ko, \href{https://doi.org/10.1007/s41365-023-01231-1}{Nucl. Sci. Tech. {\bf 34}, 80 (2023)}.

\bibitem{SunKJ} Kai-Jia Sun, Rui Wang, Che Ming Ko, Yu-Gang Ma, Chun Shen, \href{https://doi.org/10.1038/s41467-024-45474-x}{Nature Commun. {\bf 15}, 1074 (2024)}

\bibitem{Bleicher}M. Bleicher, \href{https://doi.org/10.1007/s41365-024-01477-3}{Nucl. Sci.  Tech. {\bf 35}, 129 (2024).}

\bibitem{MaYG}
Y. G. Ma, \href{https://doi.org/10.1140/epja/i2006-10119-4}{Eur. Phys. J. A {\bf 30}, 227 (2006).}

\bibitem{STAR}
M. I. Abdulhamid {\it et al.} (The STAR Collaboration),
\href{https://doi.org/10.1016/j.physletb.2023.138165}{Phys. Lett. B {\bf 845},  138165 (2023)}.

\bibitem{Bleicher2}
M. Bleicher, E. Bratkovskaya, \href{https://doi.org/10.1016/j.ppnp.2021.103920}{Prog. Part. Nucl. Phys. {\bf 122}, 103920 (2022)}.

\bibitem{CE48} C. E. Shannon, The Mathematical Theory of Communication, The University of Illinois Press, Urbana, Ill, USA, 1948.


\bibitem{YGM99} Y. G. Ma, 
              \href{https://doi.org/10.1103/PhysRevLett.83.3617}{Phys. Rev. Lett.  \textbf{83}, 3617 (1999)}.

\bibitem{HP08} H. Petersen, J. Steinheimer, G. Burau, M. Bleicher, H. St\"ocker,  \href{https://doi.org/10.1103/PhysRevC.78.044901}{Phys. Rev. C {\bf 78}, 044901 (2008)}.

\bibitem{SR80} S. R. De Groot, W. A. Van Leeuwen, and C. G. Van Weert, Relativistic Kinetic theory:\ Principles and Applications (North-Holland, Amsterdam, 1980).
                        
\bibitem{SA98} S. A. Bass, M. Belkacem, M. Bleicher {\it et al.}, 
                        \href{https://doi.org/10.1016/S0146-6410(98)00058-1}{Prog. Part. Nucl. Phys. {\bf 41}, 225 (1998)}.   
                              
\bibitem{MB99} M. Bleicher,  E. Zabrodin, C. Spieles {\it et al.}, 
                        \href{https://doi.org/10.1088/0954-3899/25/9/308}{J. Phys. G {\bf 25},  1859  (1999)}.       
                                            
\bibitem{JS18} J. Steinheimer, V. Vovchenko,  J. Aichelin {\it et al.}, \href{https://doi.org/10.1051/epjconf/201817105003}{EPJ Web of Conferences {\bf 171}, 05003 (2018)}.

\bibitem{QiaoFH} F. H. Qiao, X. G. Deng, Y. G. Ma, \href{https://doi.org/10.1016/j.physletb.2024.138535}{Phys. Lett. B {\bf 850}, 138535 (2024).}       

\bibitem{MM09} M. Mitrovski, T. Schuster, G. Gr{\"a}f  {\it et al.}, 
                      \href{https://doi.org/10.1103/PhysRevC.79.044901}{Phys. Rev. C {\bf 79}, 044901 (2009)}. 
                                            
\bibitem{PPB10} P. P. Bhaduri and S. Chattopadhyay,  
                       \href{https://doi.org/10.1103/PhysRevC.81.034906}{Phys. Rev. C {\bf 81}, 034906 (2010) }. 
                       
\bibitem{SS19} S. Sombun, K. Tomuang, A. Limphirat   {\it et al.}, 
                      \href{https://doi.org/10.1103/PhysRevC.99.014901}{Phys. Rev. C {\bf 99}, 014901 (2019)}.  
\bibitem{LiQF_NT}
 Zepeng Gao,  Qingfeng Li, \href{https://doi.org/10.11889/j.0253-3219.2023.hjs.46.080009}{Nucl. Tech. {\bf 46}, 080009 (2023)}.


\bibitem{LiQF_NT2}
Bo Gao, Yongjia Wang, Qingfeng Li, Baochun Li, 
\href{https://doi.org/10.11889/j.0253-3219.2023.hjs.46.070501}{Nucl. Tech. {\bf 46}, 070501 (2023)}.


\bibitem{LiQF_NST}
Kui Xiao, Pengcheng Li, Yongjia Wang, Fuhu Liu, Qingfeng Li, 
\href{https://doi.org/10.1007/s41365-023-01205-3}{Nucl. Sci. Tech. {\bf 34}, 62 (2023)}. 


\bibitem{SSS_NST}
S. W. Lan, S. S. Shi, \href{https://doi.org/10.1007/s41365-022-01006-0}{Nucl. Sci. Tech.  {\bf 33}, 21 (2022)}. 

\bibitem{Li_SCPMA}
P. C. Li, Y. J. Wang, Q. F. Li, H. F. Zhang, \href{https://doi.org/10.1007/s11433-022-2026-5}{Sci. Chin. Phys. Mech. and Astro. {\bf 66}, 222011 (2023)}.     

\bibitem{DZ02} D. Zschiesche, S. Schramm, J. Schaffner-Bielich, {\it et al.},
                      \href{https://doi.org/10.1016/S0370-2693(02)02736-3}{Phys. Lett. B {\bf 547}, 7 (2002)}.

\bibitem{JS11} J. Steinheimer, S. Schramm and H. St{\"o}cker, 
                      \href{https://doi.org/10.1103/PhysRevC.84.045208}{Phys. Rev. C {\bf 84}, 045208 (2011)}                      

\bibitem{DH95} D. H. Rischke, Y. P{\"u}rs{\"u}n and J. A. Maruhn, 
                      \href{https://doi.org/10.1016/0375-9474(95)00356-3}{Nucl. Phys. A {\bf 595}, 383 (1995)}.

\bibitem{YB16}Yu. B. Ivanov and A. A. Soldatov,
              \href{https://doi.org/10.1140/epja/i2016-16367-7}{Eur. Phys. J. A {\bf52}: 367 (2016)}.
              
 \bibitem{LI2009111} Q. F. Li,  J. Steinheimer, H.  Petersen, {\it et al.},
             \href{https://doi.org/10.1016/j.physletb.2009.03.012}{Phys. Lett. B {\bf 674}, 111 (2009)}.

                       
\bibitem{LH20} Hui Liu, Dingwei Zhang, Shu He, Kai-jia Sun, Ning Yu, Xiaofeng Luo,        \href{https://doi.org/10.1016/j.physletb.2020.135452}{Phys. Lett. B {\bf 805}, 135452 (2020)}.   



\bibitem{CWM18}
              Chun-Wang Ma, Yu-Gang Ma, \href{https://doi.org/10.1016/j.ppnp.2018.01.002}
             {Prog. Part. Nucl. Phys. {\bf 99}, 120  (2018).}   
             
\bibitem{LF20} F. Li and G. Chen, 
             \href{https://doi.org/10.1140/epja/s10050-020-00169-x}{Eur. Phys. J. A {\bf 56}:167 (2020)}.  

\bibitem{PuJie23} J. Pu, Y. B. Yu, K. X. Kai {\it et al.}, \href{https://doi.org/10.1016/j.physletb.2023.137909}{Phys. Lett. B {\bf 841}, 137909 (2023)}.   

\bibitem{PRD}
K. Kashiwa, H. Kouno, \href{https://doi.org/10.1103/PhysRevD.105.054017}{Phys. Rev. D {\bf 105}, 054017 (2022)}.

\bibitem{PLB}
W. Kou, X. Chen, \href{https://doi.org/10.1016/j.physletb.2023.138199}{Phys. Lett. {\bf B 846},138199 (2023)}.

\bibitem{WeiHL}
H. L. Wei, X. Zhu, C.  Yuan, \href{https://doi.org/10.1007/s41365-022-01096-w}{Nucl. Sci. Tech. {\bf 33}, 111 (2022)}.

\bibitem{Tsallis}C. Tsallis, 
\href{https://doi.org/10.1007/BF01016429}{J. Stat. Phys. {\bf 52}, 479-487 (1988).} 

\bibitem{Renyi}
P. Jizba, T. Arimitsu, \href{https://doi.org/10.1016/j.aop.2004.01.002}{Annals of Physics {\bf 312}, 17-59 (2004)}.

 \bibitem{KM15} K. Morita, B. Friman, K. Redlich,
             \href{http://dx.doi.org/10.1016/j.physletb.2014.12.037}{Phys. Lett. B {\bf 714}, 178 (2015)}.   
\bibitem{CH16} C. Herold {\it et al.},   
             \href{https://doi.org/10.1103/PhysRevC.93.021902}{ Phys. Rev. C {\bf 93}, 021902(R) (2016)}.               

             
\bibitem{STAR1_2021}M. S. Abdallah {\it et al.} (STAR Collaboration),
            \href{https://doi.org/10.1103/PhysRevC.104.024902}{Phys. Rev. C {\bf 104}, 024902 (2021)}.




\end{thebibliography}
\end{document}